# Mutual-Information-Dependent Nonlinear Threshold Response Model Linked to the Free Energy Principle

Tatsuaki Tsuruyama, M.D., Ph.D.
Department of Drug Discovery Medicine,
Graduate School of Medicine, Kyoto University, Kyoto 606-8501, Japan
Running title: Mutual Information–Dependent Threshold Responses
†Correspondence: (email:tsuruyam@kuhp.kyoto-u.ac.jp)

**Abstract**
Biological systems not only infer states of the external world from sensory input but also vary the expression of their responses according to the information they have acquired. Here, without altering the standard inferential and policy-evaluation schemes of the Free Energy Principle (FEP) and active inference, we propose a minimal dynamics that links the mutual information formed between an external state and an internal representation through inference to a response-expression variable distinct from policy selection. In this formulation, established information is positioned as a state signal that modulates response expression. The model introduces a piecewise nonlinear term in which response activation is driven only when mutual information exceeds an information threshold. Monte Carlo simulations using a two-state Markov environment showed that increasing observation accuracy increased the mutual information between the external state and the internal representation and, in turn, increased response activation. By contrast, in a control condition without information-response coupling, the response remained at its baseline level. The basic pattern of response activation was preserved when observation accuracy, the information-response coupling coefficient, the information threshold, and the closed-loop coefficient linking response to sensory sampling were varied. These results show that coupling information formed by FEP-consistent inference to an independent response-expression dynamics can generate a nonlinear response that depends on an information threshold.

## 1. Introduction

The Free Energy Principle (FEP) is a theoretical framework that describes perception, learning, and action in terms of biological systems or agents forming internal models of the external causes of sensory input and minimizing variational free energy [1,2]. Variational free energy in this context is not thermodynamic free energy itself; it is a statistical and variational quantity that evaluates the consistency between an approximate posterior distribution and a generative model. In active inference, future policies are evaluated using expected free energy, incorporating not only the attainment of preferences but also information acquisition that reduces uncertainty [3-8,16,17].

However, what an organism infers about the external world, or which policy it selects, is not identical to how strongly the selected response is expressed. Even for the same response content, movement frequency or speed, neural response gain, and intracellular signaling activity may vary continuously. To make this distinction explicit, we define a dimensionless variable normalized to the interval 0–1, termed normalized response activation, and denote it by $A_t$, which is an abstract state variable describing the degree to which a selected response is expressed.

Friston discussed the formal relationship between the FEP and the information bottleneck [20], and Sengupta et al. explicitly treated mutual information between sensory states and internal representations [21]. In active inference, expected information gain from future observations also appears in policy evaluation as epistemic value [3,7]. The question addressed here is therefore not whether mutual information can be introduced into the FEP, but what dynamical consequences arise when information that has already been established through inference is assigned a different role from future information seeking—namely, that of a state signal modulating response expression. For this purpose, we introduce a minimal model in which mutual information between the external state and the internal representation is measured and drives $A_t$ only when it exceeds a threshold. We then analyze fixed points and stability and examine the properties and robustness of the model by numerical simulation in a two-state stochastic environment, parameter sweeps, estimator-sensitivity analysis, and an active-sensing closed-loop analysis.

## 2. Results

### 2.1 Mapping Established Information to Response Expression

Let x denote a hidden state, i.e., a state that cannot be fully observed directly and is inferred as a cause of sensory observations; let y denote an observation, $p_\theta(x, y)$ the generative model, and $q(x)$the recognition distribution. For a single observation, variational free energy is defined as

$$F[q; y] = \mathbb{E}_{q(x)}[\ln q(x) - \ln p_\theta(x, y)] \quad (1)$$

is defined. Factoring the generative model as$p_\theta(x, y) = p_\theta(x|y)p_\theta(y)$, we obtain

$$F[q; y] = D_{\mathrm{KL}}\big(q(x) \parallel p_\theta(x \mid y)\big) - \ln p_\theta(y) \quad (2)$$

Therefore,

$$F[q; y] \geq -\ln p_\theta(y) \quad (3)$$

A complete derivation and a proof of the non-negativity of the KL divergence are provided in Appendix A.1. As Equation (2) shows, the KL divergence is already part of the inferential structure of the FEP [1,2]. Under a policy π, let $q(x, y|\pi)$ denote the predictive joint distribution over a future state X and future observation Y. Averaging the KL divergence between the predictive distribution before observation, q(x|π), and the predictive posterior after observation, $q(x|y, \pi)$, over future observations gives

$$\mathrm{EIG}(\pi) = \mathbb{E}_{q(y|\pi)}\big[D_{\mathrm{KL}}\big(q(x \mid y, \pi) \parallel q(x \mid \pi)\big)\big] \quad (4)$$

This quantity is equal to

$$\mathrm{EIG}(\pi) = I_q(X; Y \mid \pi) \quad (5)$$

(See **Appendix A.2**). Thus, expressing epistemic value in active inference as mutual information or expected information gain is already contained in existing theory [3,7]. The symbols are listed in **Table 1**.

### 2.2 Beliefs and Internal Representations

To simplify the analysis, we consider a minimal numerical model. The belief about a simple two-state environment is defined as

$$q_t = P(X_t = 1 \mid Y_{1:t}) \qquad (6)$$

In this minimal model, the binary internal representation is defined as

$$R_t = \begin{cases} 1, & q_t \geq \frac{1}{2}, \\ 0, & q_t < \frac{1}{2}, \end{cases} \qquad (7)$$

Because the posterior distribution can be calculated exactly in the two-state model, we use the exact Bayesian posterior corresponding to the optimum of variational free-energy minimization in the FEP. From the most recent W steps up to time *t*, we construct the empirical joint distribution $\hat{p}_t(x, r)$. The plug-in estimate of mutual information is

$$I_t = \sum_{x,r} \hat{p}_t(x, r) \ln \frac{\hat{p}_t(x,r)}{\hat{p}_t(x)\hat{p}_t(r)} \qquad (8)$$

For the ideal population distribution, $I(X; R) \geq 0$ and equals 0 when *X* and *R* are independent (See **Appendix A.3).** Equation (8) is not assumed to be explicitly computed by the biological system itself. In the simulation, it is an analyst-level information measure that can be computed because the analyst has access to the true $X_t$.

### 2. 3 Normalized Response Activation and Projection

We introduce the following discrete-time dynamics.

$$A_{t+1} = \Pi_{[0,1]}\left(A_t + \frac{A_0 - A_t}{\tau_A} + \alpha g(I_t - I_c)\right) \qquad (9)$$

Here,

$$g(z) = \max(0, z), \qquad \Pi_{[0,1]}(z) = \min\{1, \max(0, z)\} (10)$$

The first equation in (10) is a rectifying function that replaces negative inputs by 0, and $\Pi_{[0,1]}$ is the projection onto the interval [0,1]. Thus, a calculated value below 0 is set to 0, a value above 1 is set to 1, and all other values are left unchanged. This projection is a boundary condition that keeps $A_t$ within the normalized response range; the threshold nonlinearity itself is provided by $g(I_t - I_c)$. Equation (9) has three components. The first term, $A_t$, is the current response state; the second term, $(A_0 - A_t)/\tau_A$, relaxes the response toward the baseline $A_0$ in the absence of information-driven input; and the third term, $\alpha g(I_t - I_c)$, acts only when information exceeds the threshold. The coefficient α represents the effective sensitivity of $A_t$, per time step, to information above threshold, whereas $I_c$ is the threshold at which information-dependent drive begins. Neither is a thermodynamic coefficient; both are phenomenological response parameters in the present model.

$$I_t \le I_c \quad \Rightarrow \quad g(I_t - I_c) = 0 \qquad (11)$$

and there is no information-dependent drive. In contrast,

$$I_t > I_c \quad \Rightarrow \quad g(I_t - I_c) = I_t - I_c > 0 \qquad (12)$$

adds a positive drive.

### 2.4 Fixed Point under Constant Information Input

Consider the interior region in which $I_t = I$ is constant and the projection is inactive. Equation (9) becomes

$$A_{t+1} = \left(1 - \frac{1}{\tau_A}\right) A_t + \frac{A_0}{\tau_A} + \alpha g(I - I_c) \qquad (13)$$

The fixed point $A_{\mathrm{fp}}$ is

$$A_{fp} = A_0 + \tau_A \alpha g(I - I_c) \quad (14)$$

(Appendix A.4). Thus, for $I \le I_c, A_{fp} = A_0$, whereas for $I > I_c$ the fixed point begins to increase as a function of information. This kink is the minimal nonlinearity of the model.

### 2.5 Stability of the Fixed Point

Let $\delta_t = A_t - A_{\mathrm{fp}}$ denote the deviation from the fixed point. Then

$$\delta_{t+1} = \left(1 - \frac{1}{\tau_A}\right)\delta_t \qquad (15)$$

Therefore,

$$\left|1 - \frac{1}{\tau_A}\right| < 1 \quad \Leftrightarrow \quad \tau_A > \frac{1}{2} (16)$$

the interior fixed point is stable. If $r = 1 - 1/\tau_A$, then

$$A_t = A_{\text{fp}} + r^t(A_{\text{init}} - A_{\text{fp}}), \qquad r = 1 - \frac{1}{\tau_A} \; (17)$$

Let $A_th$ denote the operational response criterion. When $A_init < A_th < A_{fp}$, the continuous-valued boundary for the time required to reach the criterion is

$$T_0 = \frac{\ln\left[\frac{A_{\text{fp}} - A_{\text{th}}}{A_{\text{fp}} - A_{\text{init}}}\right]}{\ln r} \; (18)$$

In discrete time, the arrival time is the smallest integer time not less than this value (Appendix A.5). Equations (14) and (18) therefore predict, before numerical simulation, that increasing α produces a larger and faster response, whereas increasing $I_c$ makes activation more difficult and slower.

### 2.6 Numerical Model: Two-State Markov Environment

The external state is

$$X_t \in \{0,1\} \qquad (19)$$

with

$$P(X_t \neq X_{t-1}) = \varepsilon, \qquad P(X_t = X_{t-1}) = 1 - \varepsilon \qquad (20)$$

In the present analysis, ε = 0.03. The mean persistence time of one state is therefore 1/ε = 33.3 steps, and an average of 7.5 state reversals is expected within each 250-step evaluation period. This provides a minimal setting in which the environment does not remain fixed and multiple state transitions occur within the time window. The observation $Y_t$ matches the true state with probability $\eta_t$.

$$P(Y_t = X_t) = \eta_t, \qquad P(Y_t \neq X_t) = 1 - \eta_t \quad (21)$$

In the basic simulation, η = 0.55 in the first half and η = 0.90 in the second half. A value of 0.55 represents a low-information condition close to the chance level of 0.5 for binary discrimination, whereas 0.90 represents a highly discriminable condition. For an equiprobable binary symmetric channel, the ideal mutual information in the observation itself is approximately 0.005 nat at η = 0.55 and approximately 0.368 nat at η = 0.90, placing the information threshold $I_c$ = 0.18 nat between these values. However, the quantity that drives activation in this study, $I(X;R)$, is the empirical mutual information with the internal representation after Bayesian filtering and is not identical to this ideal channel value.

Let $q_t = P(X_t = 1|Y_{1:t})$. The predictive prior before observing $Y_t$ is

$$\tilde{q}_t = (1-\varepsilon)q_{t-1} + \varepsilon(1-q_{t-1}) \quad (22)$$

For the observation $Y_t = y_t$, define the likelihoods as

$$L_1 = P(Y_t = y_t \mid X_t = 1), \qquad L_0 = P(Y_t = y_t \mid X_t = 0) \quad (23)$$

Then

$$q_t = \frac{L_1\tilde{q}_t}{L_1\tilde{q}_t + L_0(1-\tilde{q}_t)} \quad (24)$$

updates the belief. We used the exact Bayesian update available for the two-state model. This allows the two stages, from observation discriminability to mutual information, and from mutual information to response activation, to be examined separately.

### 2.7 Numerical Simulation Results

In the basic simulation, increasing observation accuracy did not immediately alter the response. Instead, it first increased the accuracy of the internal representation and the mutual information, after which response activation increased when the information threshold was exceeded. The parameters are listed in **Table 2**.

### 2.7.1 Mutual information

Across 1,000 trials under the baseline condition, $I(X;R)$ was 0.0559±0.0029 nat during the pre-change evaluation interval (t = 150–239) and 0.2964±0.0061 nat during the post-change interval (t = 350–499). The accuracy of the internal representation also increased from 0.6088±0.0102 to 0.9508±0.0014 (**Figure 1**).

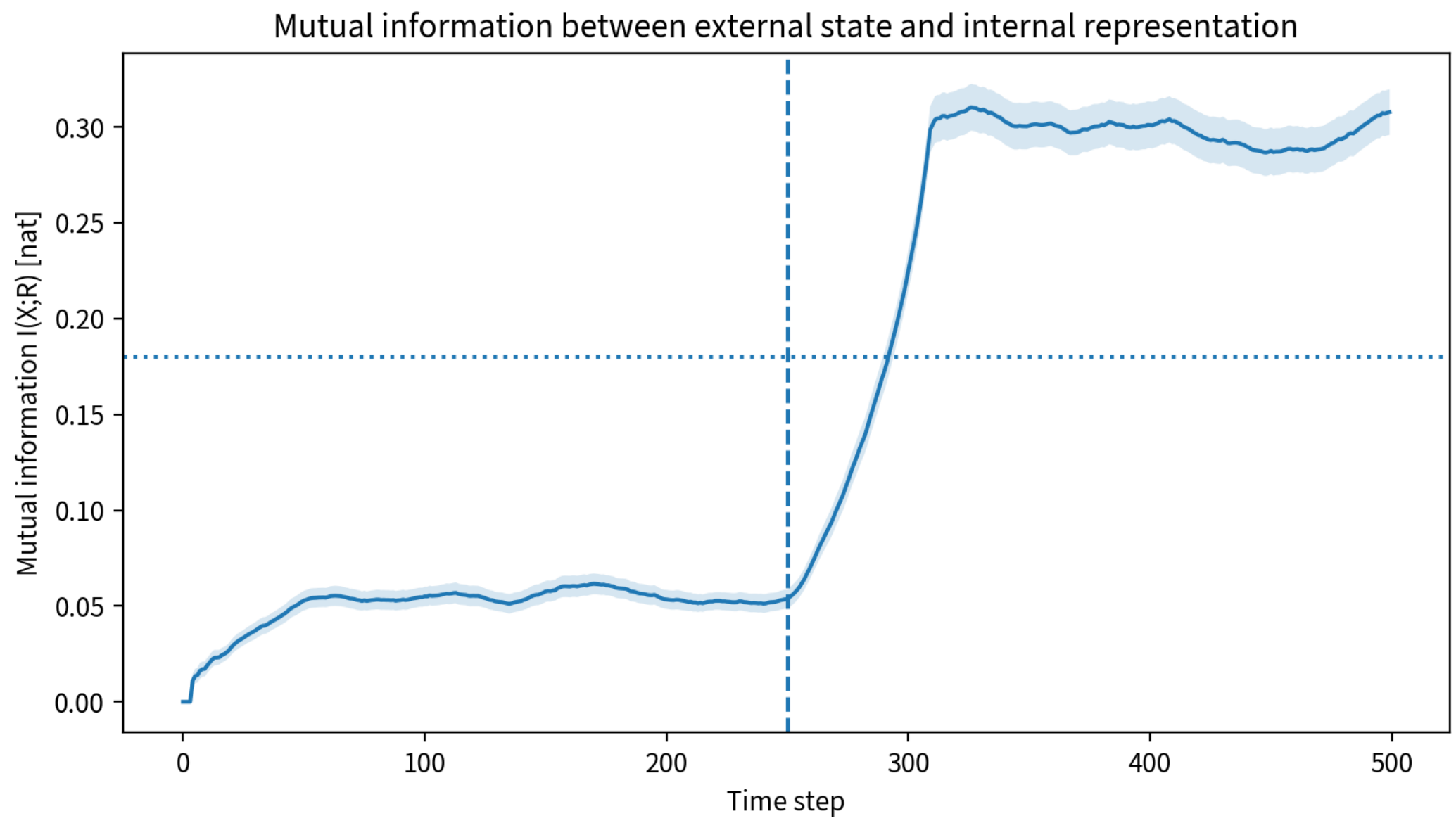

**Figure 1.** Mutual information between the external state and the internal representation. The solid line shows the mean across 1,000 trials and the shaded band shows the 95% confidence interval across trials. The vertical dashed line marks the time at which observation accuracy was changed from 0.55 to 0.90; the horizontal dotted line indicates $I_c$ = 0.18 nat.

This result shows that activation was not increased directly from outside the model; rather, increased observation accuracy first improved inference and thereby increased $I(X;R)$.

### 2.7.2 Information–Response Coupling

In the same trials, normalized response activation $A_t$ increased from 0.1860±0.0047 in the first half to 0.6978±0.0114 in the second half. When α = 0 was used as a control, the external state, observations,

Bayesian filter, and therefore the process of internal-representation formation were unchanged, but the information-driven term vanished and $A_t$ converged to $A_0 = 0.15$. This control shows that improved inference alone does not produce response activation; a coupling that maps established information onto the response is required (**Figure 2**).

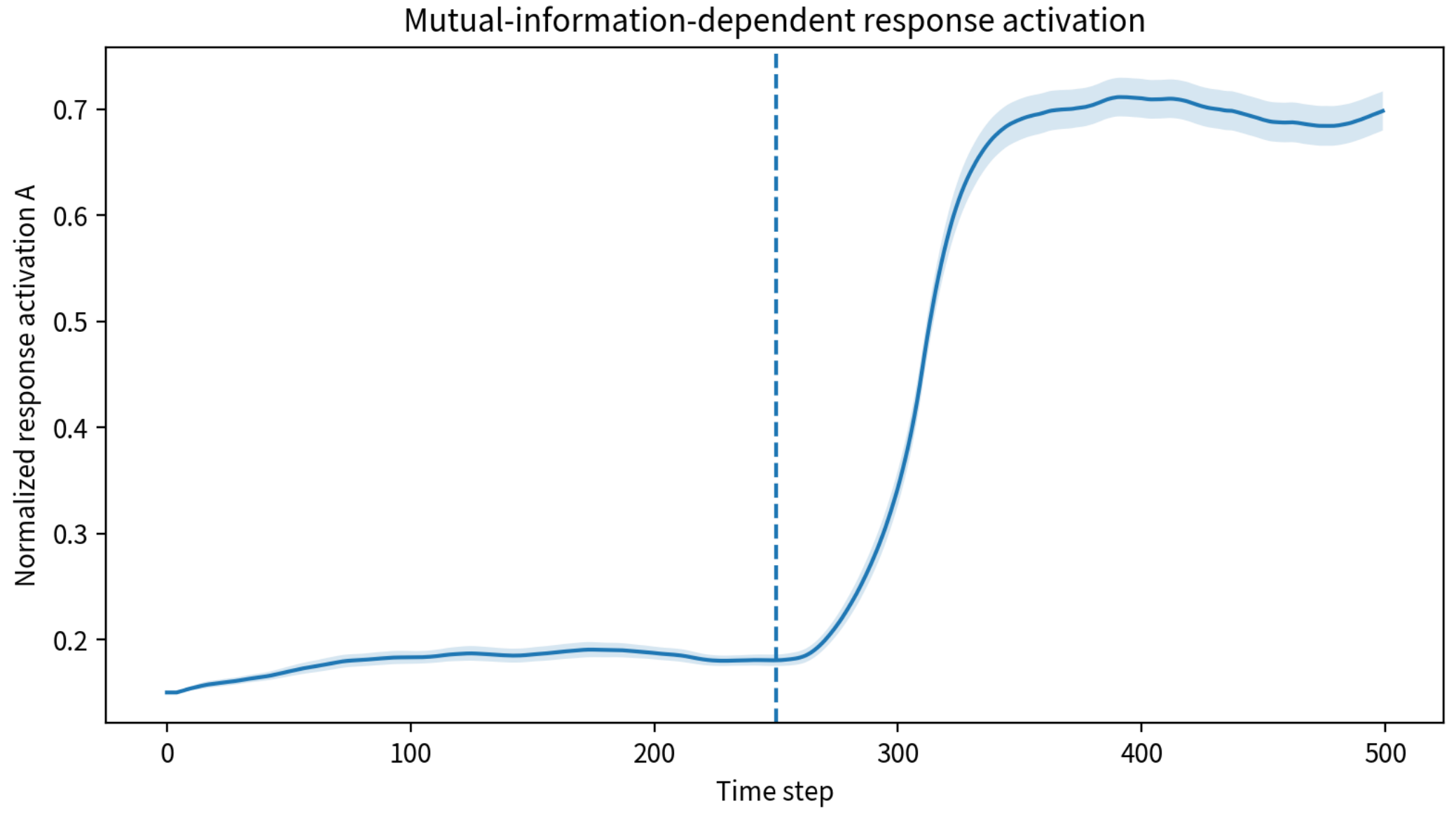

**Figure 2**. Mutual-information-dependent normalized response activation. The solid line shows the mean across 1,000 trials and the shaded band shows the 95% confidence interval across trials.

Thus, in the present model, established information is converted into a change in response only when the mapping from mutual information to response activation in Equation (9) is present.

### 2.7.3 Observation Accuracy

We next varied observation accuracy η from 0.50 to 0.95 in increments of 0.025, with 500 trials for each condition. The steady-state means changed as follows. At η = 0.75, mean $I(X;R) = 0.180$ nat, approximately equal to $I_c$, while mean $A$ remained 0.430. At η = 0.80, mean $I(X;R) = 0.206$ nat and mean $A$ increased to 0.505 (**Figure 3**). Thus, the response is not simply proportional to observation accuracy $\eta$. Instead, increasing observation accuracy improves the internal representation, which raises $I(X;R)$ above $I_c$ and thereby activates the response. Finite-window estimation and trial-to-trial variation smooth the population average, but the individual update rule has an explicit kink at $I_c$.

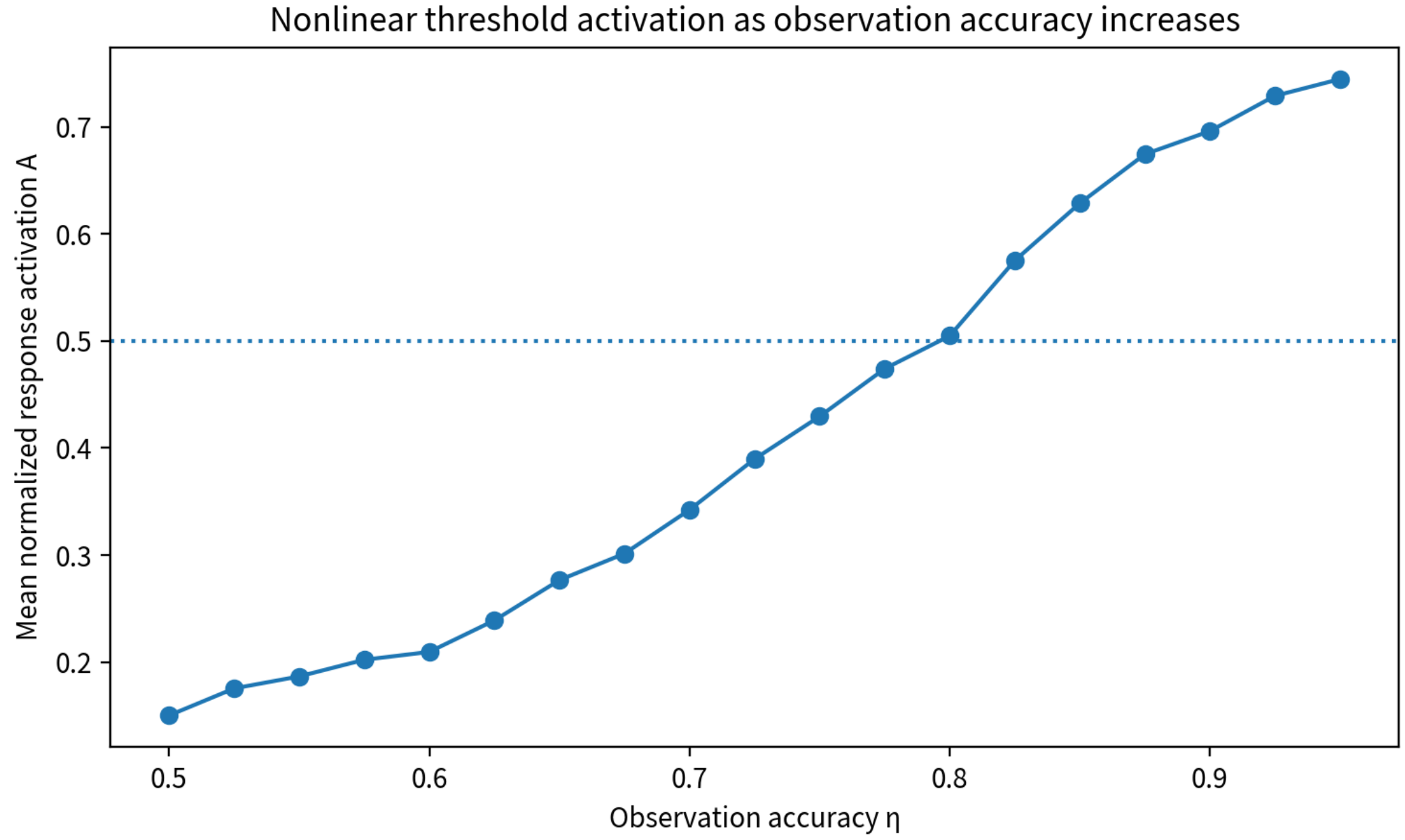

Figure 3. Mean normalized response activation as a function of observation accuracy η. The horizontal dotted line indicates the operational criterion A = 0.5. The response is not simply proportional to $\eta$; it rises as the information threshold is crossed.

### 2.7.4 α–$I_c$ Parameter Sweep

We swept α from 0.00 to 0.30 and $I_c$ from 0.05 to 0.35 nat in increments of 0.01, using the same 500 information trajectories for all parameter combinations. With α fixed at 0.18, $I_c$ = 0.05 yielded a post-change mean $A$ of 0.815, a response-arrival fraction of 0.998, and a mean latency of 50.4 steps. Under the baseline condition $I_c$ = 0.18, the corresponding values were 0.694, 0.994, and 76.1 steps, whereas at $I_c$=0.35 they were 0.394, 0.824, and 115.6 steps. As predicted by Equations (14) and (18), higher thresholds produced smaller and slower responses (**Figure 4**).

Figure 4 specifically summarizes response magnitude, whereas the response-arrival fraction and latency measures quoted above are shown in Figures 5 and 6, respectively. Across the full α–$I_c$ parameter space, mean post-change response activation increased with α and decreased as $I_c$ was raised (Figure 4). The fraction of trials reaching $A$≥0.5 was low at small α and increased with α, with larger $I_c$ requiring larger α for frequent criterion attainment (**Figure 5**). Among conditions that reached the criterion, mean latency decreased with increasing α and increased with $I_c$; conditions in which the criterion was never reached are shown as missing values (**Figure 6**). Thus, Figures 4–6 summarize response magnitude, response occurrence, and response timing, respectively, over the same parameter sweep.

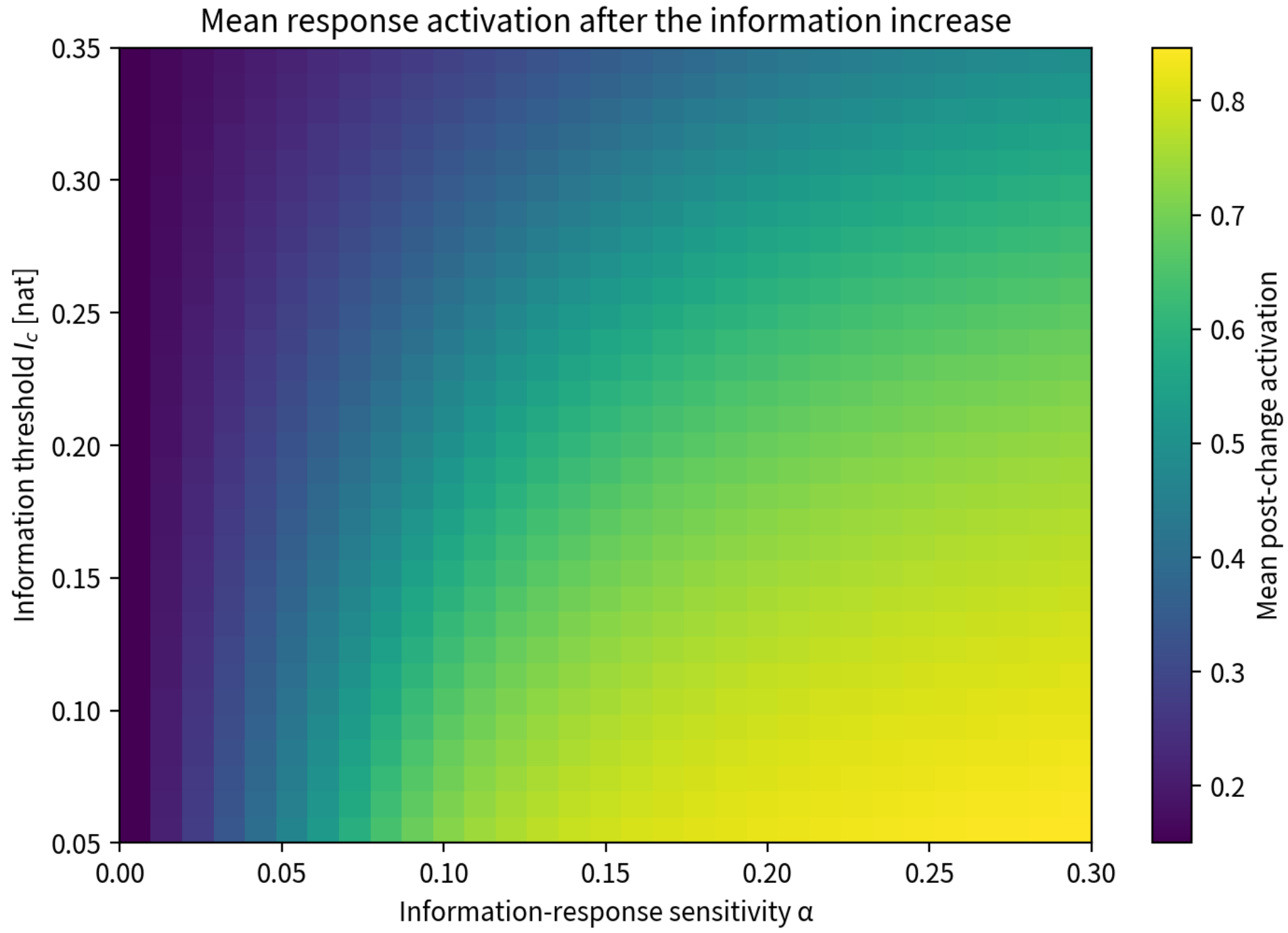

**Figure 4.** Mean normalized response activation after the information increase across the α–$I_c$ parameter space.

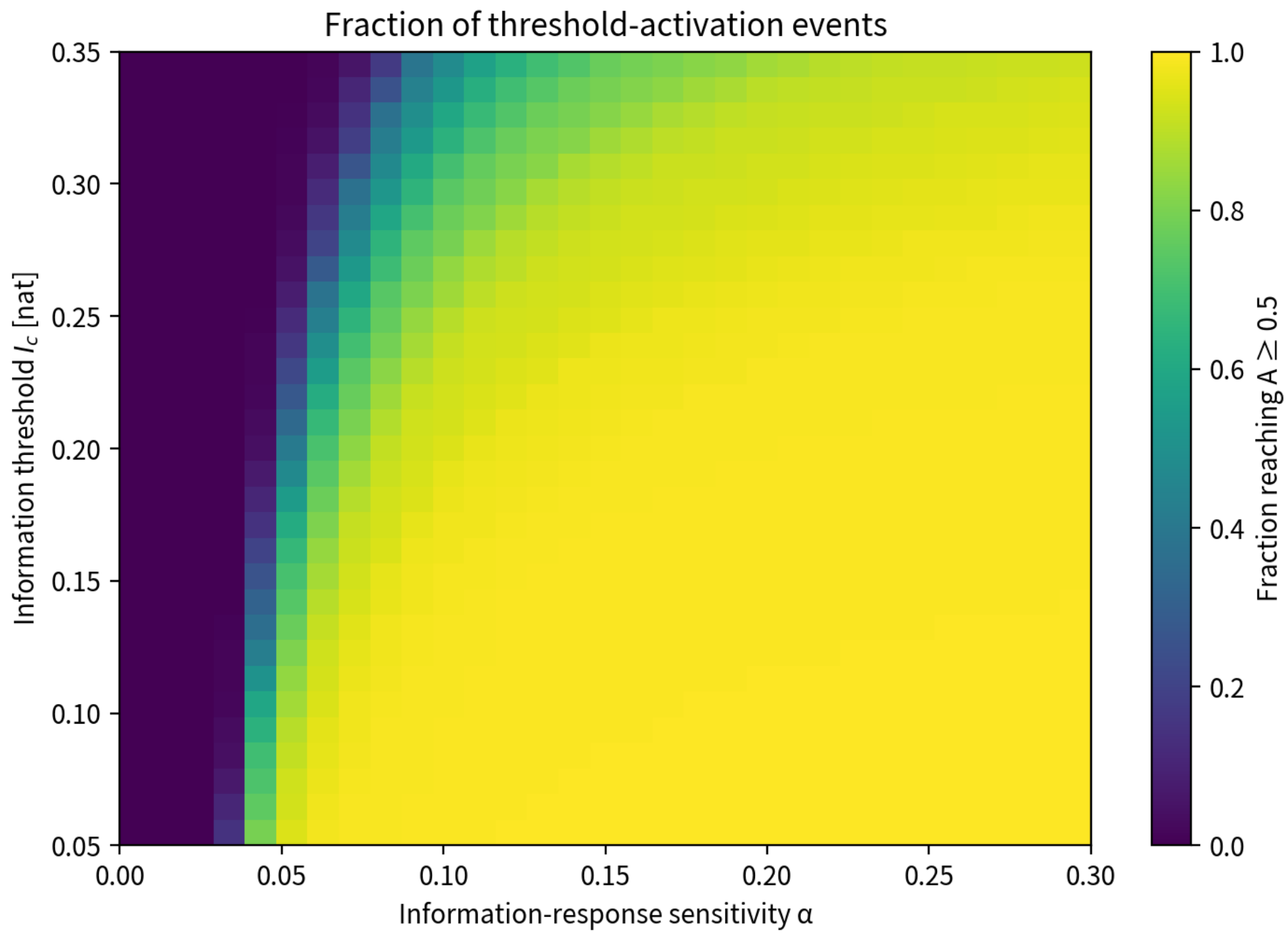

**Figure 5.** Fraction of trials reaching the operational criterion A≥0.5. This is not a phase boundary of a physical phase transition, but the frequency of threshold activation in a finite set of trials.

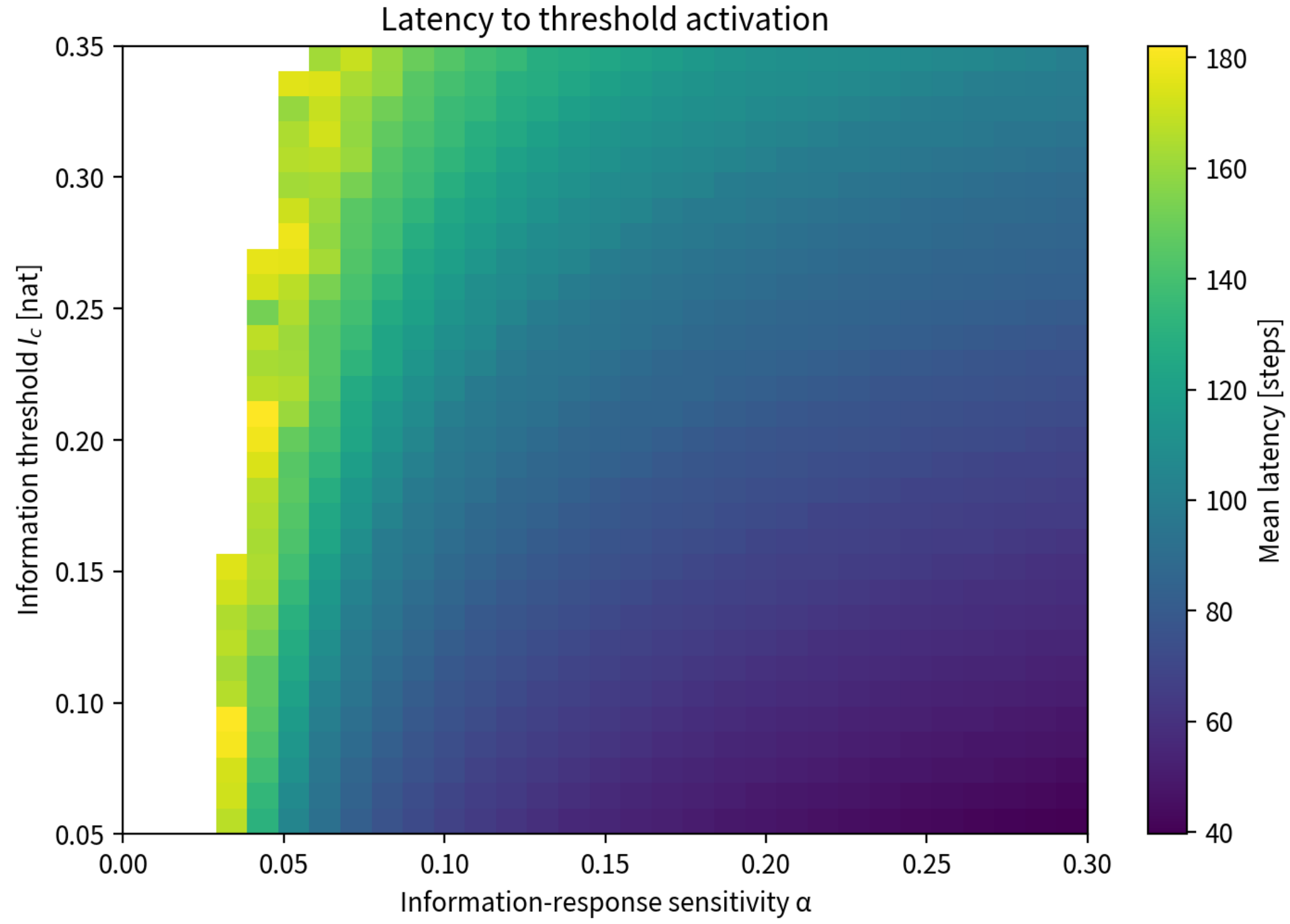

**Figure 6.** Mean latency from the change in observation accuracy to the first attainment of A≥0.5. Conditions in which the criterion was never reached are shown as missing values.

**2.7. 5 Minimal Closed Loop for Sensory Sampling**

As a minimal model of active sensing, we define the probability of acquiring an observation using the closed-loop coefficient $\beta$ as

$$s_t = \Pi_{[0,1]}[s_0 + \beta(A_t - A_0)] \qquad (25)$$

We set this probability according to Equation (25). When β = 0, $s_t = s_0$ and response activation does not feed back to sensory sampling. When β>0, higher response activation makes acquisition of a new observation more likely. The closed-loop analysis used $s_0$ = 0.35 and 800 trials for each β (**Figure 7**).

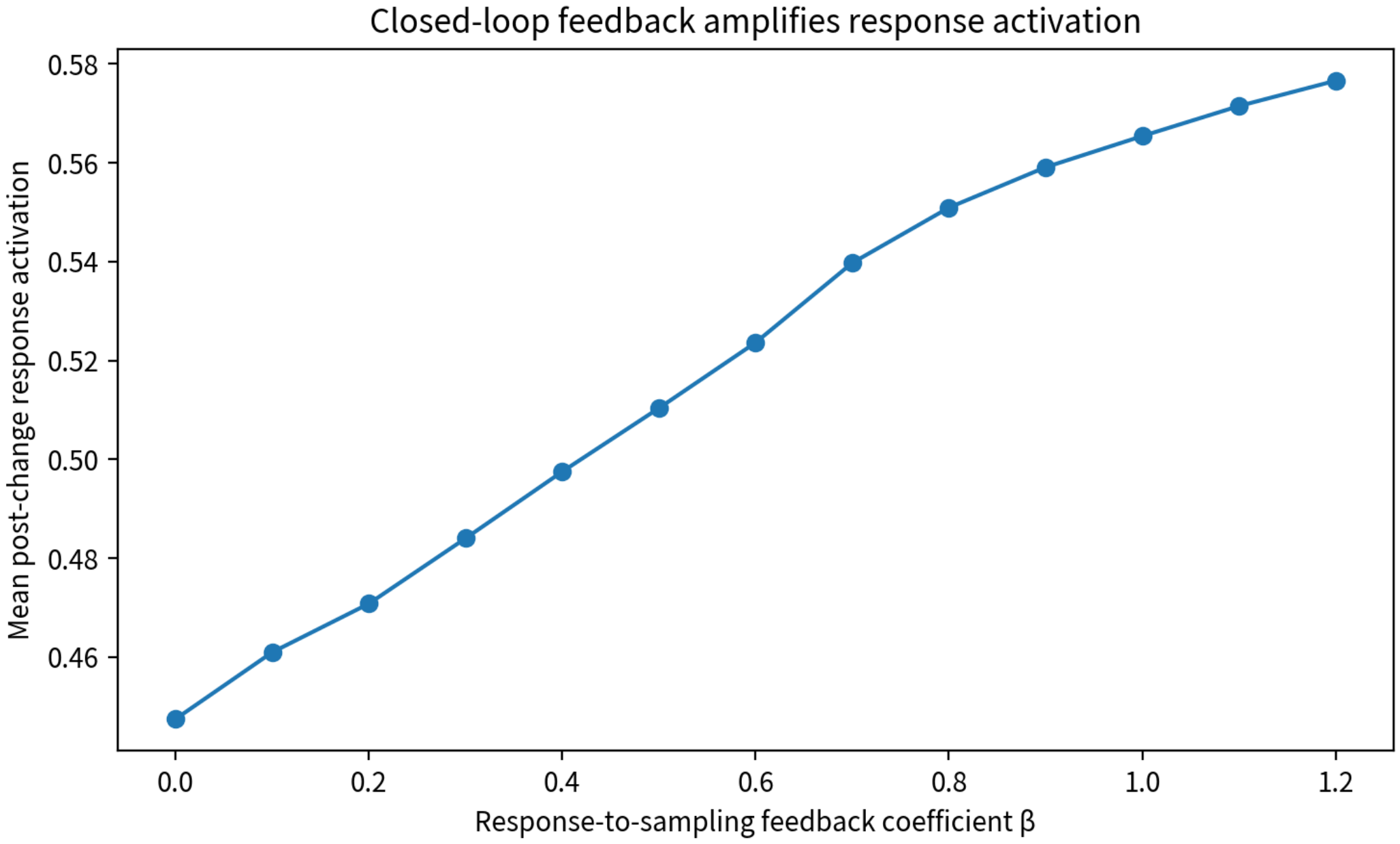

**Figure 7.** Feedback coefficient β from response activation to observation acquisition and mean post-change response activation. Stronger positive feedback systematically amplifies the response.

For β>0, increased response activation raised the probability of acquiring the next observation, and those observations in turn improved internal-representation accuracy and mutual information, producing positive feedback. As $\beta$ increased, post-change mean A, $I(X;R)$, observation-acquisition probability, and internal-representation accuracy all increased systematically. Thus, the one-way information to response coupling was not lost when extended to an active-sensing closed loop; rather, it could be amplified.

## 3. Discussion

The use of mutual information to quantify statistical dependence between external or sensory states and internal representations has precedents in the literature. Friston [20] showed a formal relationship between the FEP and the information bottleneck, and Sengupta et al. [21] treated mutual information between sensory and internal states. In active inference, the value of future information is also expressed as expected information gain or mutual information [3,7]. In the present model, however, the focus is not the value of information to be obtained in the future. Instead, the informational correspondence $I(X;R)$ already established through inference is used as an input to an independent response-expression dynamics. This separation of roles yields the prediction that even when beliefs or policies are the same, the magnitude and time course of response expression can differ if α and $I_c$ differ.

The model yields at least three experimentally testable predictions. First, there may be conditions under which response onset is better explained by how discriminatively the internal state represents the external state than by stimulus intensity itself. Second, if a biological mechanism can alter the information-response sensitivity α, response magnitude may change while the accuracy of the internal representation remains nearly unchanged. Third, if a mechanism can alter the information threshold $I_c$, the condition for initiating a response should shift systematically for the same amount of information. The proposed information-response coupling can therefore be tested experimentally by measuring the environmental state, decoding accuracy of the internal representation, and response expression simultaneously.

Several simplifications were made to isolate the minimal structure of information-response coupling. First, because the analyst knows the true $X_t$ in the simulation, $I(X;R)$ can be calculated directly; real biological systems, however, do not have direct access to the true hidden state. Application to a specific biological system will therefore require an internally available proxy correlated with $I(X;R)$, such as posterior confidence, prediction-error statistics, receptor states, or neural population states. Second, reducing $R_t$ to a binary MAP representation discards uncertainty contained in the posterior distribution. Future extensions should consider$I(X;Z)$ for continuous internal states$Z_t$or sufficient statistics of the posterior distribution.

Third, the closed loop considered here includes only the pathway by which the response changes the probability of sensory acquisition. In systems such as chemotaxis or foraging, where responses alter the environment itself, an environment-mediated loop of the form $A_t \rightarrow X_{t+1} \rightarrow Y_{t+1} \rightarrow q_{t+1} \rightarrow I_{t+1} \rightarrow A_{t+2}$ must be formulated for the biological system of interest. In addition, the current threshold nonlinearity arises from the explicitly introduced rectifying function. To examine genuine bifurcations or multistability, future models should introduce response-generating systems with Hill functions, sigmoids, or positive feedback and analyze the number of fixed points, bifurcation diagrams, hysteresis, and noise-induced transition rates. Accordingly, the present claim is limited to threshold-induced nonlinear responses in a finite system.

We have situated the relationship between the FEP and mutual information within existing work and proposed a nonlinear threshold dynamics that links mutual information established through inference to normalized response activation. The central proposal is not to add mutual information to the FEP objective, but to map established information onto an independent response-expression variable. The analysis showed that when information remains at or below $I_c$, only the baseline response is stable, whereas above $I_c$ the fixed-point increases with α and the amount of information exceeding the threshold. Numerical analysis further showed a temporal sequence in which increased observation accuracy improved the internal representation and mutual information before response activation emerged. This basic structure was preserved when observation accuracy, α, $I_c$, mutual-information estimation conditions, and the closed-loop coefficient β were varied. Thus, when FEP-consistent inference forms an internal representation that retains sufficient information about the environment, coupling that established information to a response-expression dynamics distinct from policy selection can produce a threshold-induced nonlinear response. This result suggests a theoretical role for information not only in determining what is inferred or which action is selected, but also in regulating how strongly a selected response is expressed. The model provides a minimal framework for experimentally testing the relationship between information retained in internal representations and response expression in neurobehavioral systems, cellular signaling systems, and active-sensing systems.

## Declaration

### Consent to participate

Not applicable.

### Consent to publish

Not applicable. This manuscript does not contain identifiable images or data of participants.

### Funding

This research received no external funding.

### Data availability

Not applicable. This manuscript does not report data generation or analysis.

### Ethics declaration

The author declares that the study does not involve humans nor animal subjects.

## Tables

### Table 1 Symbol list

| Symbol | Definition |
|---|---|
| $X_t$ | External state (hidden state) at time t |
| $Y_t$ | Observation at time t |
| $q_t$ | $\left(P(X_t = 1 \mid Y_{1:t})\right)$: posterior belief that the state is 1 |
| $R_t$ | Binary internal representation: $(R_t = 1) if (q_t \geq 1/2), and (R_t = 0)$ otherwise |
| $\varepsilon$ | One-step switching probability of the external state |
| $\eta_t$ | Probability that observation $Y_t$ matches $X_t$ (observation accuracy) |
| $W$ | Moving-window length used to estimate mutual information |
| $I_t$ | $I(X;R)$estimated from the most recent W time steps |
| $A_t$ | Normalized response activity |
| $A_0$ | Baseline response activity in the absence of information-driven input |
| $\tau_A$ | Relaxation time constant governing the return of $A_t$ toward $A_0$ |
| $\alpha$ | Sensitivity coefficient linking information to response activity |
| $I_c$ | Mutual-information threshold for the onset of information-dependent responses |
| $s_t, s_0$ | Observation acquisition probability in the closed loop and its baseline value, respectively |
| $\beta$ | Feedback coefficient from response activity to observation acquisition |

**Table 2. Baseline simulation parameters and rationale for their selection**

| Parameter | Baseline value | Rationale for selection |
|---|---|---|
| T | 500 steps | Provides 250 steps before and after the change for a steady-state comparison. |
| Monte Carlo trials | 1000 | Stabilizes trial-averaged estimates and 95% confidence intervals. |
| ε | 0.03 / step | Gives a mean dwell time of 33.3 steps and multiple state reversals in each half. |
| η | 0.55 → 0.90 | Compares a low-information condition near chance with a highly discriminable condition. |
| Change time | t = 250 | Makes the pre- and post-change periods equal in length. |
| W | 60 steps | About 1.8 times the mean dwell time, balancing sample size and temporal resolution. |
| $A_0$ | 0.15 | Illustrative baseline response within the 0–1 interval. |
| $\tau_A$ | 25 steps | Shorter than W, allowing tracking of MI changes while avoiding an instantaneous response. |
| $\alpha$ | 0.18 | Illustrative value that avoids immediate saturation at A = 1 under the high-information condition. |
| $I_c$ | 0.18 nat | Intermediate threshold separating the low- and high-information conditions. |
| Arrival criterion | A≥0.50 | Operational criterion used to compare event fractions and latencies. |

## Appendix A. Mathematical Derivations and Proofs

### A.1 Variational Free-Energy Decomposition and the Surprisal Bound

Starting from Equation (1),

$$F[q;y] = \int q(x)[\ln q(x) - \ln p_\theta(x,y)]dx$$

Bayes' product rule gives

$$p_\theta(x,y) = p_\theta(x \mid y)p_\theta(y)$$

Substituting this expression yields

$$F[q;y] = \int q(x)[\ln q(x) - \ln p_\theta(x \mid y) - \ln p_\theta(y)]dx$$

Because -log p_θ(y) does not depend on x,

$$F[q;y] = \int q(x)\ln\frac{q(x)}{p_\theta(x \mid y)}dx - \ln p_\theta(y)\int q(x)dx$$

$$\int q(x)dx = 1$$

$$F[q;y] = D_{\mathrm{KL}}\big(q(x) \parallel p_\theta(x \mid y)\big) - \ln p_\theta(y)$$

We next show that the KL divergence is non-negative. For u>0, log u≤u-1. Setting u = p/q gives

$$-\ln\frac{p}{q} \geq 1 - \frac{p}{q}$$

Multiplying both sides by q and integrating gives

$$\int q\ln\frac{q}{p} \geq \int (q-p) = 0$$

Thus, the KL divergence is non-negative and Equation (3) follows. Equality holds only when the two distributions coincide almost everywhere.

### A.2 Expected Information Gain and Conditional Mutual Information

Substituting the definition of KL divergence into Equation (4) gives

$$\mathrm{EIG}(\pi) = \sum_y q(y \mid \pi)\sum_x q(x \mid y,\pi)\ln\frac{q(x \mid y,\pi)}{q(x \mid \pi)}$$

Using the product rule q(x,y|π) = q(y|π)q(x|y,π),

$$\mathrm{EIG}(\pi) = \sum_{x,y} q(x,y \mid \pi)\ln\frac{q(x \mid y,\pi)}{q(x \mid \pi)}$$

Furthermore,

$$q(x \mid y, \pi) = \frac{q(x, y \mid \pi)}{q(y \mid \pi)}$$

and therefore

$$\mathrm{EIG}(\pi) = \sum_{x,y} q\,(x, y \mid \pi) \ln \frac{q(x, y \mid \pi)}{q(x \mid \pi) q(y \mid \pi)} = I_q(X; Y \mid \pi)$$

### A.3 Non-Negativity of Mutual Information

Mutual information is

$$I(X; R) = \sum_{x,r} p\,(x, r) \ln \frac{p(x, r)}{p(x) p(r)}$$

and

$$I(X; R) = D_{\mathrm{KL}}\big(p(X, R) \parallel p(X) p(R)\big)$$

By the non-negativity of the KL divergence shown in A.1, I(X;R)≥0. Moreover, I(X;R) = 0 if and only if

$$p(x, r) = p(x) p(r)$$

which is equivalent to independence of X and R.

### A.4 Fixed Point and Stability of the Response-Activation Equation

Assume $I_t$ = I is constant and that the projection is inactive. The fixed point of Equation (13) satisfies

$$A_{\mathrm{fp}} = \left(1 - \frac{1}{\tau_A}\right) A_{\mathrm{fp}} + \frac{A_0}{\tau_A} + \alpha g(I - I_c)$$

Subtracting $(1 - 1/\tau_A) A_{fp}$ from both sides gives

$$\frac{A_{\mathrm{fp}}}{\tau_A} = \frac{A_0}{\tau_A} + \alpha g(I - I_c)$$

and therefore

$$A_{\mathrm{fp}} = A_0 + \tau_A \alpha g(I - I_c)$$

Next, let δ_t denote a small deviation from the fixed point. Subtracting the fixed-point equation gives

$$\delta_{t+1} = \left(1 - \frac{1}{\tau_A}\right) \delta_t$$

A first-order discrete-time system converges if and only if the absolute value of its update coefficient is less than 1. Therefore, Equation (16) shows that the interior fixed point is stable when the relaxation time constant is greater than 1/2.

In the region where information input exceeds the threshold,

$$\frac{\partial A_{\mathrm{fp}}}{\partial \alpha} = \tau_A (I - I_c) > 0, \qquad \frac{\partial A_{\mathrm{fp}}}{\partial I_c} = -\tau_A \alpha < 0$$

Hence, increasing the information-response coupling coefficient α raises the fixed point, whereas increasing the information threshold I_c lowers it.

A.5 Time to Reach the Operational Threshold

Iterating Equation (15) gives

$$\delta_t = r^t \delta_0, \qquad r = 1 - \frac{1}{\tau_A}$$

and therefore

$$A_t = A_{\mathrm{fp}} + r^t \left(A_{\mathrm{init}} - A_{\mathrm{fp}}\right)$$

At the boundary at which the operational response criterion is reached with equality,

$$A_{\mathrm{fp}} - A_{\mathrm{th}} = r^t \left(A_{\mathrm{fp}} - A_{\mathrm{init}}\right)$$

Normalizing both sides by the positive difference gives

$$r^t = \frac{A_{\mathrm{fp}} - A_{\mathrm{th}}}{A_{\mathrm{fp}} - A_{\mathrm{init}}}$$

Taking the natural logarithm of both sides yields Equation (18). If the fixed point is at or below the operational threshold, the criterion is not reached as long as the trajectory approaches the fixed point monotonically.

A.6 Local Stability of the Closed-Loop Mean-Field Approximation

In the closed loop, approximate the long-time average mutual information as a smooth function 𝒥(A) of response activation. On the activated branch 𝒥(A)>I_c,

$$A_{t+1} = A_t + \frac{A_0 - A_t}{\tau_A} + \alpha[\mathcal{J}(A_t) - I_c]$$

The fixed point is

$$A_{\mathrm{fp}} = A_0 + \tau_A \alpha\left[\mathcal{J}\left(A_{\mathrm{fp}}\right) - I_c\right] \qquad \text{(A1)}$$

Introduce a small deviation δ_t around the fixed point and write

$$\mathcal{J}(A_t) \approx \mathcal{J}\left(A_{\mathrm{fp}}\right) + \mathcal{J}'\left(A_{\mathrm{fp}}\right)\delta_t$$

to first order. Using the fixed-point condition gives

$$\delta_{t+1} = \left[1 - \frac{1}{\tau_A} + \alpha \mathcal{J}'\left(A_{\mathrm{fp}}\right)\right]\delta_t$$

Therefore, the local stability condition is

$$\left|1-\frac{1}{\tau_A}+\alpha\mathcal{J}'(A_{\mathrm{fp}})\right|<1 \qquad \text{(A2)}$$

Assume positive feedback in which increasing response activation increases information acquisition, so that the slope of mean information with respect to activation is non-negative. If the relaxation time constant is greater than 1/2, the upper stability condition then gives

$$\tau_A\alpha\mathcal{J}'(A_{\mathrm{fp}})<1 \qquad \text{(A3)}$$

The left-hand side can be interpreted as a closed-loop gain L. As L approaches 1, amplification by positive feedback becomes stronger; when L exceeds 1, this interior fixed point becomes unstable under the linear approximation. This is only a candidate condition for bifurcation and does not establish a thermodynamic phase transition in the present simulations.

## Appendix B. Numerical Results and Reproducibility Conditions

### B.1 Basic Simulation

The basic simulation used T = 500, 1,000 trials, and seed 20260809. The pre-change evaluation interval was 150–239 and the post-change interval was 350–499. The results are reported in the main text. In the control condition α = 0, $A_t$converges to $A_0$ = 0.15.

### B.2 Parameter Sweep

α = 0.00–0.30 and $I_c$ = 0.05–0.35 nat were both evaluated in increments of 0.01. The same 500 external-state, observation, and inference trajectories were shared across all conditions. The criterion A≥0.5 used in the main text is an operational comparison criterion, not a physical phase boundary.

B.3 Closed-Loop Analysis

We used s_0 = 0.35 and β = 0.0–1.2 in increments of 0.1, with 800 trials per condition and seed 20260812. Random-number sequences used for external-state reversals, observation correctness, and observation-acquisition decisions were shared across β conditions.

## Appendix C. Reproduction Python Code

The following code reproduces the basic simulation, observation-accuracy sweep, estimator-sensitivity analysis, $\alpha - I_c$ sweep, and closed-loop β sweep reported in the main text. Mutual information is calculated using natural logarithms and reported in nats. The function projection01 corresponds to $\Pi_{[0,1]}$ in the text; the ambiguous programming term clip is not used.

```python
"""Reproduction code for the FEP-linked mutual-information threshold activation model.
Python 3.13.5; NumPy 2.3.5; pandas 2.2.3; Matplotlib 3.10.8.
All logarithms are natural logarithms; mutual information is reported in nats.
"""
import numpy as np
import pandas as pd
import matplotlib.pyplot as plt

plt.rcParams["font.family"] = "Noto Sans CJK JP"
plt.rcParams["axes.unicode_minus"] = False

# ---- Core parameters ----
T = 500
P_SWITCH = 0.03
ETA_LOW, ETA_HIGH = 0.55, 0.90
CHANGE_T = 250
W = 60
A0 = 0.15
TAU_A = 25.0
ALPHA = 0.18
I_C = 0.18
A_TRANS = 0.50
SEED_BASIC = 20260809
SEED_ETA = 20260811
SEED_CLOSED = 20260812


def projection01(z):
    """Projection Pi_[0,1](z) = min(1,max(0,z))."""
    return np.minimum(1.0, np.maximum(0.0, z))
```

```python
def rolling_binary_mi(X, R, w = W, correction = None):
    """Rolling plug-in mutual information I(X;R), in nats, for binary variables."""
    nrep, n_time = X.shape
    out = np.zeros((nrep, n_time), dtype = float)
    for t in range(n_time):
        s = max(0, t - w + 1)
        xx, rr = X[:, s:t+1], R[:, s:t+1]
        n = xx.shape[1]
        if n < 5:
            continue
        c00 = np.sum((xx == 0) & (rr == 0), axis = 1)
        c01 = np.sum((xx == 0) & (rr == 1), axis = 1)
        c10 = np.sum((xx == 1) & (rr == 0), axis = 1)
        c11 = n - c00 - c01 - c10
        counts = np.stack([c00, c01, c10, c11], axis = 1).reshape(-1, 2, 2)
        pxy = counts / n
        px = pxy.sum(axis = 2, keepdims = True)
        pr = pxy.sum(axis = 1, keepdims = True)
        denom = px * pr
        term = np.zeros_like(pxy, dtype = float)
        mask = (pxy > 0) & (denom > 0)
        term[mask] = pxy[mask] * np.log(pxy[mask] / denom[mask])
        mi = term.sum(axis = (1, 2))
        if correction == "miller_madow":
            kx = (px[:, :, 0] > 0).sum(axis = 1)
            kr = (pr[:, 0, :] > 0).sum(axis = 1)
            mi = np.maximum(0.0, mi - ((kx - 1) * (kr - 1)) / (2 * n))
        out[:, t] = mi
    return out


def generate_inference(nrep, eta_schedule, seed):
    """Generate a two-state Markov environment and exact Bayesian filtering."""
```

```
    rng = np.random.default_rng(seed)
    X = np.zeros((nrep, T), dtype = np.int8)
    q = np.zeros((nrep, T), dtype = float)
    R = np.zeros((nrep, T), dtype = np.int8)
    X[:, 0] = rng.integers(0, 2, size = nrep)
    u_flip = rng.random((nrep, T))
    u_correct = rng.random((nrep, T))
    for t in range(T):
        if t > 0:
            flip = u_flip[:, t] < P_SWITCH
            X[:, t] = np.where(flip, 1 - X[:, t-1], X[:, t-1])
        eta = float(eta_schedule[t])
        y = np.where(u_correct[:, t] < eta, X[:, t], 1 - X[:, t])
        if t = = 0:
            prior = np.full(nrep, 0.5)
        else:
            prior = q[:, t-1] * (1 - P_SWITCH) + (1 - q[:, t-1]) * P_SWITCH
        l1 = np.where(y = = 1, eta, 1 - eta)
        l0 = np.where(y = = 1, 1 - eta, eta)
        q[:, t] = l1 * prior / (l1 * prior + l0 * (1 - prior))
        R[:, t] = (q[:, t] > = 0.5).astype(np.int8)
    return X, R, q


def activity_from_info(I, alpha = ALPHA, ic = I_C):
    """Update normalized response activation A using the threshold equation."""
    nrep, n_time = I.shape
    A = np.zeros((nrep, n_time), dtype = float)
    A[:, 0] = A0
    for t in range(n_time - 1):
        drive = alpha * np.maximum(0.0, I[:, t] - ic)
        raw = A[:, t] + (A0 - A[:, t]) / TAU_A + drive
        A[:, t+1] = projection01(raw)
```

```
    return A


def basic_simulation(nrep = 1000, w = W, correction = None, seed = SEED_BASIC):
    eta = np.full(T, ETA_LOW, dtype = float)
    eta[CHANGE_T:] = ETA_HIGH
    X, R, q = generate_inference(nrep, eta, seed)
    I = rolling_binary_mi(X, R, w = w, correction = correction)
    A = activity_from_info(I)
    return X, R, q, I, A


def ci95_trial_mean(arr, start, stop):
    by_trial = arr[:, start:stop].mean(axis = 1)
    mean = by_trial.mean()
    halfwidth = 1.96 * by_trial.std(ddof = 1) / np.sqrt(len(by_trial))
    return mean, halfwidth


def ideal_binary_channel_mi(eta):
    """I(X;Y) for an equiprobable binary symmetric channel, in nats."""
    e = 1.0 - eta
    if e == 0.0 or e == 1.0:
        h = 0.0
    else:
        h = -(e * np.log(e) + (1 - e) * np.log(1 - e))
    return np.log(2.0) - h


# ---- Basic simulation ----
X, R, q, I, A = basic_simulation()
metrics = []
for name, arr, interval in [
```

```
    ("I_pre", I, (150, 240)), ("I_post", I, (350, 500)),
    ("A_pre", A, (150, 240)), ("A_post", A, (350, 500)),
    ("accuracy_pre", (X == R).astype(float), (150, 240)),
    ("accuracy_post", (X == R).astype(float), (350, 500)),
]:
    m, hw = ci95_trial_mean(arr, *interval)
    metrics.append((name, m, hw))
pd.DataFrame(metrics, columns = ["metric", "mean", "95CI_halfwidth"]).to_csv(
    "fep_basic_summary.csv", index = False
)

# ---- Observation-accuracy sweep ----
eta_rows = []
for k, eta_value in enumerate(np.arange(0.50, 0.951, 0.025)):
    eta = np.full(T, eta_value, dtype = float)
    Xe, Re, _ = generate_inference(500, eta, SEED_ETA + k)
    Ie = rolling_binary_mi(Xe, Re)
    Ae = activity_from_info(Ie)
    sl = slice(350, 500)
    eta_rows.append((
        eta_value,
        Ie[:, sl].mean(),
        Ae[:, sl].mean(),
        np.mean(Ae[:, sl].mean(axis = 1) >= A_TRANS),
        np.mean(Xe[:, sl] == Re[:, sl]),
        ideal_binary_channel_mi(eta_value),
    ))
pd.DataFrame(eta_rows, columns = [
    "eta", "mean_I_XR", "mean_A", "fraction_mean_A_ge_0.5",
    "representation_accuracy", "ideal_I_XY"
]).to_csv("fep_eta_sweep.csv", index = False)

# ---- Estimator/window sensitivity ----
```

```
sensitivity = []
for w in (40, 60, 80):
    for corr in (None, "miller_madow"):
        Xs, Rs, _, Is, As = basic_simulation(
            nrep = 500, w = w, correction = corr,
            seed = SEED_BASIC + w + (1000 if corr else 0)
        )
        sensitivity.append((
            w, "plugin" if corr is None else "Miller-Madow",
            Is[:, 150:240].mean(), Is[:, 350:500].mean(),
            As[:, 150:240].mean(), As[:, 350:500].mean(),
        ))
pd.DataFrame(sensitivity, columns = [
    "W", "estimator", "I_pre", "I_post", "A_pre", "A_post"
]).to_csv("fep_estimator_sensitivity.csv", index = False)

# ---- alpha-Ic parameter sweep with common random numbers ----
eta = np.full(T, ETA_LOW, dtype = float); eta[CHANGE_T:] = ETA_HIGH
Xp, Rp, _ = generate_inference(500, eta, SEED_BASIC)
Ip = rolling_binary_mi(Xp, Rp)
alpha_values = np.arange(0.00, 0.301, 0.01)
ic_values = np.arange(0.05, 0.351, 0.01)
sweep_rows = []
for alpha in alpha_values:
    for ic in ic_values:
        Ap = activity_from_info(Ip, alpha = alpha, ic = ic)
        by_trial_pre = Ap[:, 150:240].mean(axis = 1)
        by_trial_post = Ap[:, 350:500].mean(axis = 1)
        hits = Ap[:, CHANGE_T:] > = A_TRANS
        has_hit = hits.any(axis = 1)
        first = np.argmax(hits, axis = 1)
        latency = first[has_hit].mean() if np.any(has_hit) else np.nan
        sweep_rows.append((
```

```
                alpha, ic, by_trial_pre.mean(), by_trial_post.mean(),
                np.mean(has_hit), latency
            ))
pd.DataFrame(sweep_rows, columns = [
    "alpha", "Ic", "mean_pre_A", "mean_post_A",
    "transition_fraction", "mean_latency"
]).to_csv("fep_alpha_Ic_sweep_revised.csv", index = False)

# ---- Closed-loop active-sampling model ----
def closed_loop(beta, nrep = 800, seed = SEED_CLOSED):
    rng = np.random.default_rng(seed)
    eta = np.full(T, ETA_LOW, dtype = float); eta[CHANGE_T:] = ETA_HIGH
    X = np.zeros((nrep, T), dtype = np.int8)
    R = np.zeros((nrep, T), dtype = np.int8)
    q = np.zeros((nrep, T), dtype = float)
    I = np.zeros((nrep, T), dtype = float)
    A = np.zeros((nrep, T), dtype = float); A[:, 0] = A0
    S = np.zeros((nrep, T), dtype = float)
    X[:, 0] = rng.integers(0, 2, size = nrep)
    u_flip = rng.random((nrep, T))
    u_correct = rng.random((nrep, T))
    u_observe = rng.random((nrep, T))
    for t in range(T):
        if t > 0:
            flip = u_flip[:, t] < P_SWITCH
            X[:, t] = np.where(flip, 1 - X[:, t-1], X[:, t-1])
        e = eta[t]
        y = np.where(u_correct[:, t] < e, X[:, t], 1 - X[:, t])
        S[:, t] = projection01(0.35 + beta * (A[:, t] - A0))
        observed = u_observe[:, t] < S[:, t]
        if t = = 0:
            prior = np.full(nrep, 0.5)
        else:
```

```
            prior = q[:, t-1] * (1 - P_SWITCH) + (1 - q[:, t-1]) * P_SWITCH
        l1 = np.where(y == 1, e, 1 - e)
        l0 = np.where(y == 1, 1 - e, e)
        post = l1 * prior / (l1 * prior + l0 * (1 - prior))
        q[:, t] = np.where(observed, post, prior)
        R[:, t] = (q[:, t] >= 0.5).astype(np.int8)
        # Current rolling MI over the last W pairs only.
        start = max(0, t - W + 1)
        xx, rr = X[:, start:t+1], R[:, start:t+1]
        n = xx.shape[1]
        if n >= 5:
            c00 = np.sum((xx == 0) & (rr == 0), axis = 1)
            c01 = np.sum((xx == 0) & (rr == 1), axis = 1)
            c10 = np.sum((xx == 1) & (rr == 0), axis = 1)
            c11 = n - c00 - c01 - c10
            pxy = np.stack([c00,c01,c10,c11], axis = 1).reshape(-1,2,2) / n
            px = pxy.sum(axis = 2, keepdims = True); pr = pxy.sum(axis = 1, keepdims
= True)
            denom = px * pr; term = np.zeros_like(pxy, dtype = float)
            mask = (pxy > 0) & (denom > 0)
            term[mask] = pxy[mask] * np.log(pxy[mask] / denom[mask])
            I[:, t] = term.sum(axis = (1,2))
        if t < T - 1:
            raw = A[:, t] + (A0 - A[:, t]) / TAU_A + ALPHA * np.maximum(0.0, I[:, t] -
 I_C)
            A[:, t+1] = projection01(raw)
    return X, R, I, A, S

closed_rows = []
for beta in np.arange(0.0, 1.201, 0.1):
    Xc, Rc, Ic, Ac, Sc = closed_loop(float(beta))
    sl = slice(350, 500)
    closed_rows.append((
```

```
            beta, Ac[:, sl].mean(), Ic[:, sl].mean(),
            Sc[:, sl].mean(), np.mean(Xc[:, sl] == Rc[:, sl])
        ))
pd.DataFrame(closed_rows, columns = [
        "beta", "mean_A", "mean_I", "mean_sampling_probability",
        "representation_accuracy"
]).to_csv("fep_closed_loop_sweep.csv", index = False)

print("Reproduction completed.")
```